\documentclass[10pt,conference,letterpaper]{IEEEtran}
\usepackage[USenglish]{babel}
\usepackage{times,amsmath}
\usepackage[pdftex]{graphicx}
   \graphicspath{{./pdf/}{./png/}{./jpeg/}{./eps/}{./png_old/}}
   \DeclareGraphicsExtensions{.pdf,.png,.jpeg,.eps}
\usepackage{cite}
\usepackage{url}
\usepackage{fmtcount}
\usepackage{color}
\usepackage[tight,footnotesize]{subfigure}
\usepackage{multicol}
\usepackage{amssymb}
\usepackage{wasysym}
\usepackage{calc}
\usepackage{fix-cm}
\usepackage[activate={true,nocompatibility},final,tracking=true,factor=500,stretch=40,shrink=40]{microtype}
\def\cal{\fam2}

\newcommand{\BalphaPM}{{{\cal{B}}^{\boldmath \alpha_+}_{\boldmath \alpha_-}}}
\definecolor{DarkRed}{rgb}{0.5,0,0}
\definecolor{DarkGreen}{rgb}{0,0.5,0}
\definecolor{DarkerGreen}{rgb}{0,0.3333,0}
\definecolor{DarkBlue}{rgb}{0,0,0.75}
\definecolor{RoyalBlue}{rgb}{0,0.1373,0.4000}
\definecolor{NavyBlue}{rgb}{0,0,0.5020}
\definecolor{CobaltBlue}{rgb}{0,0.2784,0.6706}
\definecolor{lightlightgray}{rgb}{0.96875,0.96875,0.96875}
\definecolor{cyan}{rgb}{0,1,1}
\newcommand{\beginlabel}[2]{%
\begin{#1}\label{#2}}
\begin{document}
\pagestyle{plain}
\title{Bandwidth Is Not Enough:\, ``Hidden" Outlier Noise and Its Mitigation}
\author{\IEEEauthorblockN{Alexei V. Nikitin}
\IEEEauthorblockA{
Nonlinear LLC\\
Wamego, Kansas, USA\\
E-mail: avn@nonlinearcorp.com}
\and
\IEEEauthorblockN{Ruslan L. Davidchack}
\IEEEauthorblockA{Dept. of Mathematics, U. of Leicester\\
Leicester, UK\\
E-mail: rld8@leicester.ac.uk}}
\maketitle
\begin{abstract}
In addition to ever-present thermal noise, various communication and sensor systems can contain significant amounts of interference with outlier (e.g. impulsive) characteristics. Such outlier interference (including that caused by nonlinear signal distortions, e.g. clipping) can be efficiently mitigated in real-time using intermittently nonlinear filters. Depending on the interference nature and composition, improvements in the quality of the signal of interest achieved by such filtering will vary from ``no harm" to substantial. In this tutorial, we explain in detail why the underlying outlier nature of interference often remains obscured, discussing the many challenges and misconceptions associated with state-of-art analog and/or digital nonlinear mitigation techniques, especially when addressing complex practical interference scenarios. We then focus on the methodology and tools for real-time outlier noise mitigation, demonstrating how the ``excess band" observation of outlier noise enables its efficient in-band mitigation. We introduce the basic real-time nonlinear components that are used for outlier noise filtering and provide examples of their implementation. We further describe complementary nonlinear filtering arrangements for wide- and narrow-band outlier noise reduction, providing several illustrations of their performance and the effect on channel capacity. Finally, we outline ``effectively analog" digital implementations of these filtering structures, discuss their broader applications, and comment on the ongoing development of the platform for their demonstration and testing. To emphasize the effectiveness and versatility of this approach, in our examples we use particularly challenging waveforms that severely obscure low-amplitude outlier noise, such as broadband chirp signals (e.g. used in radar, sonar, and spread-spectrum communications) and ``bursty," high crest factor signals (e.g. OFDM).
\end{abstract}
\maketitle
\section*{Introduction} \label{sec:introduction}
Interfering signals originating from a multitude of natural and technogenic (man-made) phenomena often have intrinsic ``outlier" temporal and/or amplitude structures that are different from the Gaussian structure of the thermal noise. The presence of different types of such outlier noise is widely acknowledged in multiple applications, under various general and application-specific names, most commonly as {\em impulsive\/}, {\em transient\/}, {\em burst\/}, or {\em crackling\/} noise. For example, outlier electromagnetic interference (EMI) is inherent in digital electronics and communication systems, transmitted into a system in various ways, including electrostatic coupling, electromagnetic induction, or RF radiation. However, although the detrimental effects of EMI are broadly acknowledged in the industry, its outlier nature often remains indistinct, and its omnipresence and impact, and thus the potential for its mitigation, remain greatly underutilized.

There are two fundamental reasons why the outlier nature of many technogenic interference sources is often dismissed as irrelevant. The first one is a simple lack of motivation. Without using nonlinear filtering techniques the resulting signal quality is largely invariant to a particular time-amplitude makeup of the interfering signal and depends mainly on the total power and the spectral composition of the interference in the passband of interest. Thus, unless the interference results in obvious, clearly identifiable outliers in the signal's band, the ``hidden" outlier noise does not attract attention. The second reason is the highly ambiguous and elusive nature of outlier noise, and the inadequacy of tools used for its consistent observation and/or quantification. More important, the amplitude distribution of a non-Gaussian signal is generally modifiable by linear filtering, and such filtering can often convert the signal from sub-Gaussian into super-Gaussian, and {\em vice versa\/}. Thus apparent outliers in a signal can disappear and reappear due to various filtering effects, including fading and multipath, as the signal propagates through media and/or the signal processing chain.

This tutorial provides a concise overview of the methodology and tools, including their analog and digital implementations, for real-time mitigation of outlier interference in general and ``hidden" wideband outlier noise in particular. Such mitigation is performed as a ``first line of defense" against interference ahead of, or in the process of, reducing the bandwidth to that of the signal of interest. Either used by itself, or in combination with subsequent interference mitigation techniques, this approach provides interference mitigation levels otherwise unattainable, with the effects, depending on particular interference scenarios, ranging from ``no harm" to spectacular. Although the main focus of this filtering technique is mitigation of wideband outlier noise affecting a band-limited signal of interest, it can also be used, given some {\em a priori} knowledge of the signal of interest's structure, to reduce outlier interference that is confined to the signal's band.

The tutorial consists of a two-part presentation followed by a demonstration of a hardware prototype implementing the presented filtering approach. Part~I of the presentation explains in detail why underlying outlier nature of interference often remains obscured, and demonstrates how its ``excess band" observation enables its in-band mitigation. It discusses many challenges and misconceptions associated with the state-of-art analog and/or digital nonlinear mitigation techniques, especially when addressing complex practical interference scenarios.

The focus of Part~II is the methodology and tools for real-time outlier noise mitigation. First, it introduces the basic real-time nonlinear components that are used in outlier noise filtering and gives examples of their implementation. It further describes complete intermittently nonlinear filtering arrangements for wide- and narrow-band outlier noise reduction, and provides several illustrations of their performance and the effect on channel capacity. Penultimately, it outlines and illustrates ``effectively analog" digital implementation of these filtering structures, including their modifications for addressing complex practical interference scenarios. Finally, it briefly discusses broader applications of these nonlinear filtering techniques, discusses the ongoing development of the platform for their demonstration and testing, and proposes the direction of future efforts.

\section*{Part I: Many challenges and misconceptions} \label{sec:part I}
The purpose of Part~I is two-fold. First, we demonstrate that noise that appears as non-outlier (e.g. Gaussian) when observed in signal's band can in fact be ``hidden" wideband outlier (e.g. impulsive) interference. Second, we show that, while such outlier interference may be tricky to observe and even harder to track and quantify, the ability to detect the presence of such interference enables its in-band mitigation beyond levels attainable when disregarding its outlier origins.

\subsection{Outlier noise: Ubiquitous but often elusive} \label{subsec:ubiquitous}
We begin with a brief explanation of origins and omnipresence of the outlier interference, and provide several illustrations of how different coupling mechanisms and various filtering effects greatly affect its time-domain appearance.

\begin{figure}[!b]
\centering{\includegraphics[width=8.6cm]{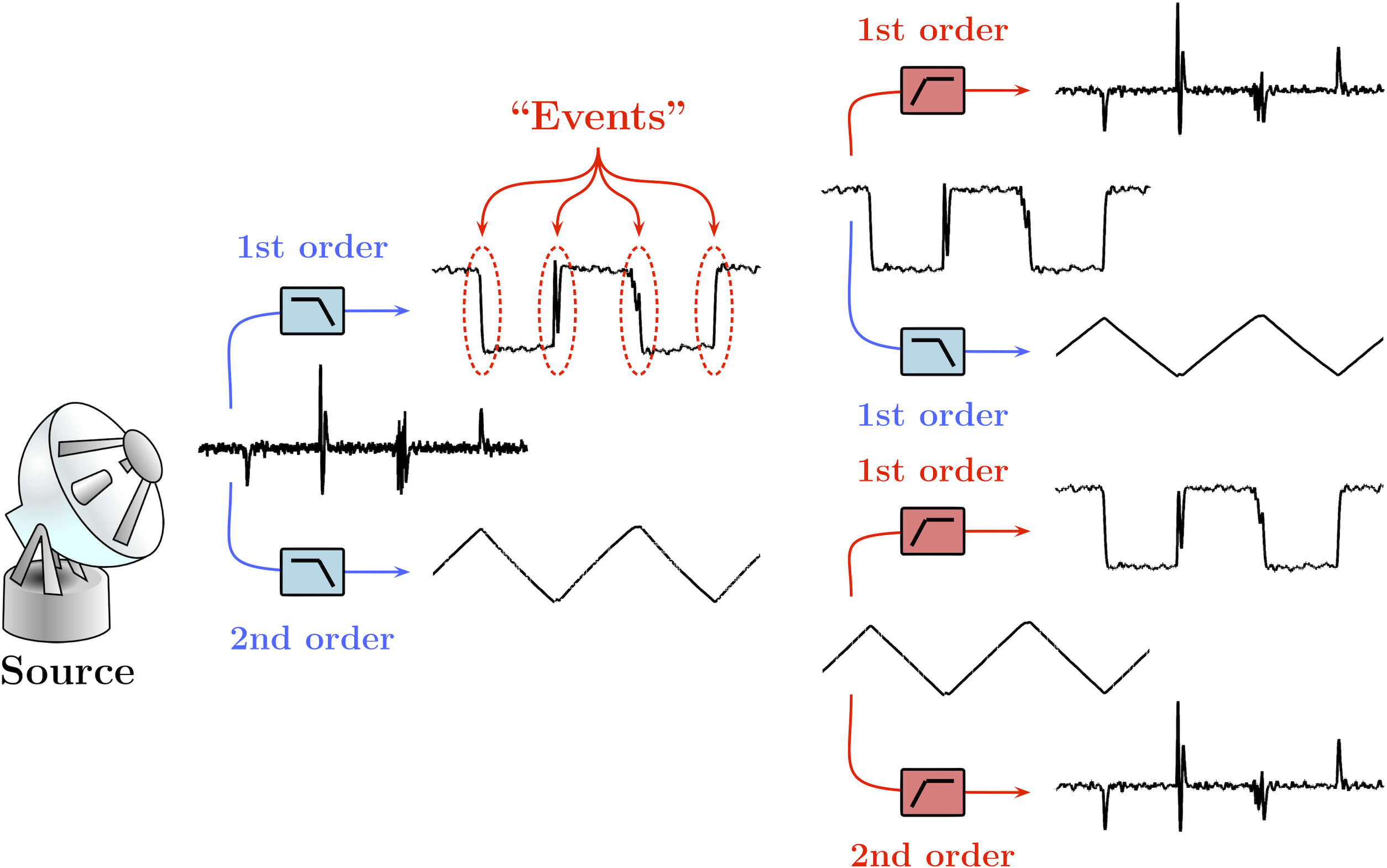}}
\caption{Outlier noise produced by ``events" separated by ``inactivity" when observed at wide bandwidth.
\label{fig:morphing}}
\end{figure}

For example, many interfering signals originating from natural and technogenic (man-made) phenomena can contain components that are produced by some ``countable" or ``discrete," relatively short duration events separated by relatively longer periods of inactivity. Given the same sequence of events, the time-domain appearance of such components can vary greatly, depending on the coupling mechanisms and the system's and propagation media's filtering properties. For instance, while all three broadband signals shown in Fig.~\ref{fig:morphing} are produced by the same sequence of events, their time domain appearances, as well as the spectral densities, are very different due to different system responses and/or filtering effects, and only one of these signals contains clearly visible time-domain outliers. Nevertheless, these signals represent the same source and they can be morphed into each other by simple 1st or 2nd order filters.

\subsection{Outlier noise: Why care? What works?} \label{subsec:why care}
We then explain and demonstrate how and why the outlier nature of interference provides an opportunity for its ``additional" in-band, real-time mitigation. Indeed, at any given frequency, a linear filter affects all signals proportionally. Thus, when linear filtering is used to suppress interference, the resulting signal quality is largely invariant to a particular makeup of the interfering signal and depends mainly on the total power and the spectral composition of the interference in the passband of interest. On the other hand, properly implemented intermittently nonlinear filtering enables in-band, real-time reduction of interference with distinct outlier components, achieving mitigation levels unattainable by linear filters~\cite{Nikitin19hidden, Nikitin18ADiC-ICC, Nikitin19ADiCpatent, Nikitin19ADiCpatentCIPs}. We further provide several ``toy" audible illustrations and experimental examples (e.g., from~\cite{Nikitin12aHSDPA} and unpublished work) of such mitigation. 

\begin{figure}[!b]
\centering{\includegraphics[width=8.6cm]{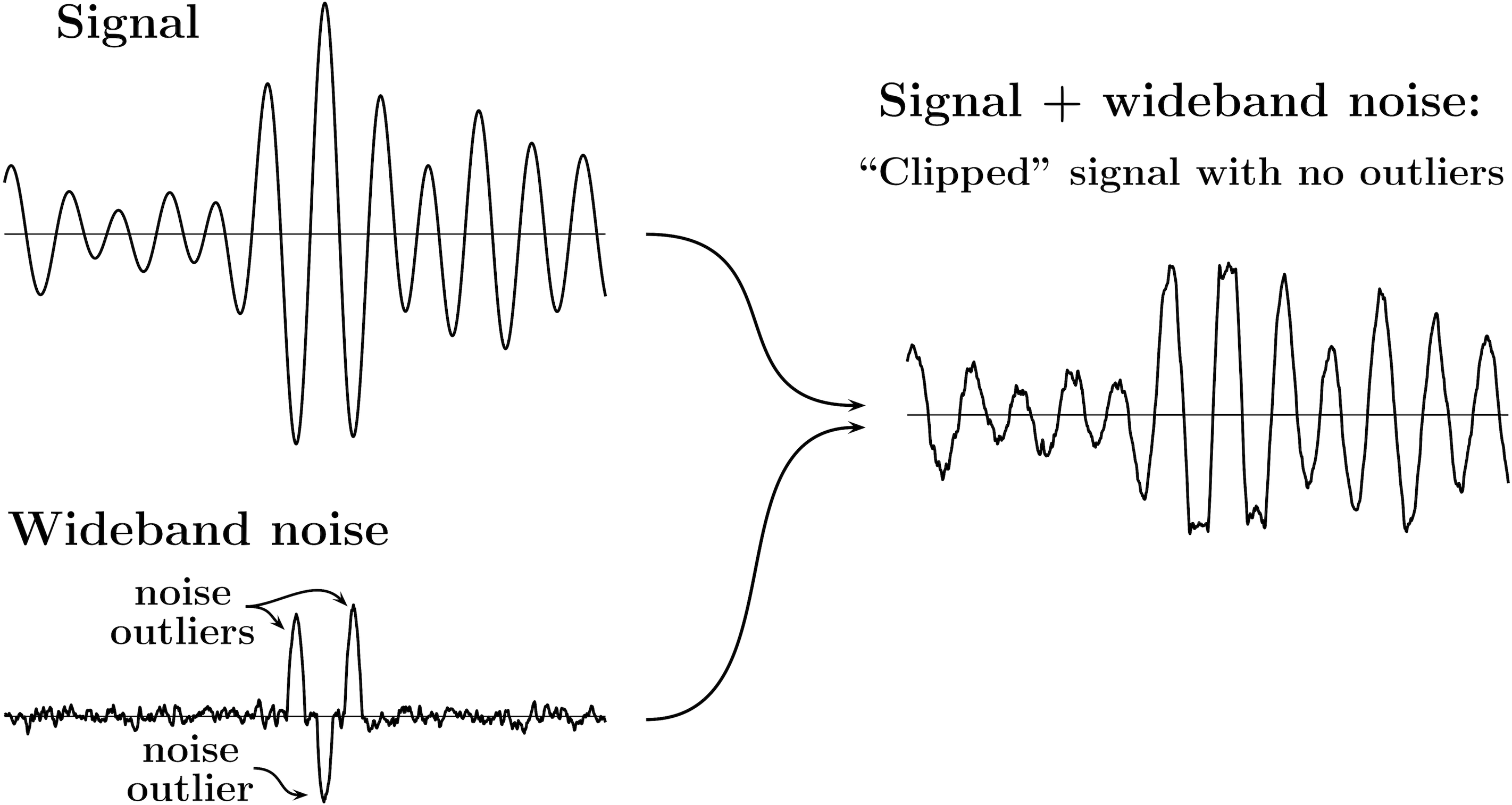}}
\caption{Interference due to clipping clearly contains outliers, yet resulting signal is outlier-free.
\label{fig:clipping}}
\end{figure}

\subsection{What hides outlier noise?} \label{subsec:what hides}
Next, we discuss several interrelated reasons why underlying outlier nature of interference often remains obscured. Those include, but are not limited to, the following:
\subsubsection{General filtering effects} \label{subsubsec:general filtering effects}
For example, as illustrated in Fig.~\ref{fig:morphing}, apparent outliers in a signal can disappear and reappear due to various filtering effects, including fading and multipath, as the signal propagates through media and/or the signal processing chain, and such filtering can make apparent outliers wax and wane. In the analog domain, such filtering can be viewed as a linear combination of the signal with its derivatives and antiderivatives (e.g. convolution) of various orders. In the digital domain, it is a combination of differencing and summation operations.

\subsubsection{``Outliers" vs ``outlier noise" ambiguity} \label{subsubsec:outlier ambiguity}
Even when the wideband noise itself contains clearly identifiable outliers, the noise outliers would not necessarily be observable as outliers in the signal+noise mixture. That would be the case, e.g.,  when the typical amplitude of the noise outliers is not significantly larger than that of the signal of interest. In those scenarios removing outliers from the signal+noise mixture may degrade the signal quality instead of improving it. As an additional example, Fig.~\ref{fig:clipping} illustrates such ``outliers" vs. ``outlier noise" ambiguity when the interference due to clipping clearly contains outliers, yet the resulting signal is outlier-free.

\subsubsection{Insufficient observation bandwidth} \label{subsubsec:insufficient bandwidth}
\begin{figure}[!b]
\centering{\includegraphics[width=8.6cm]{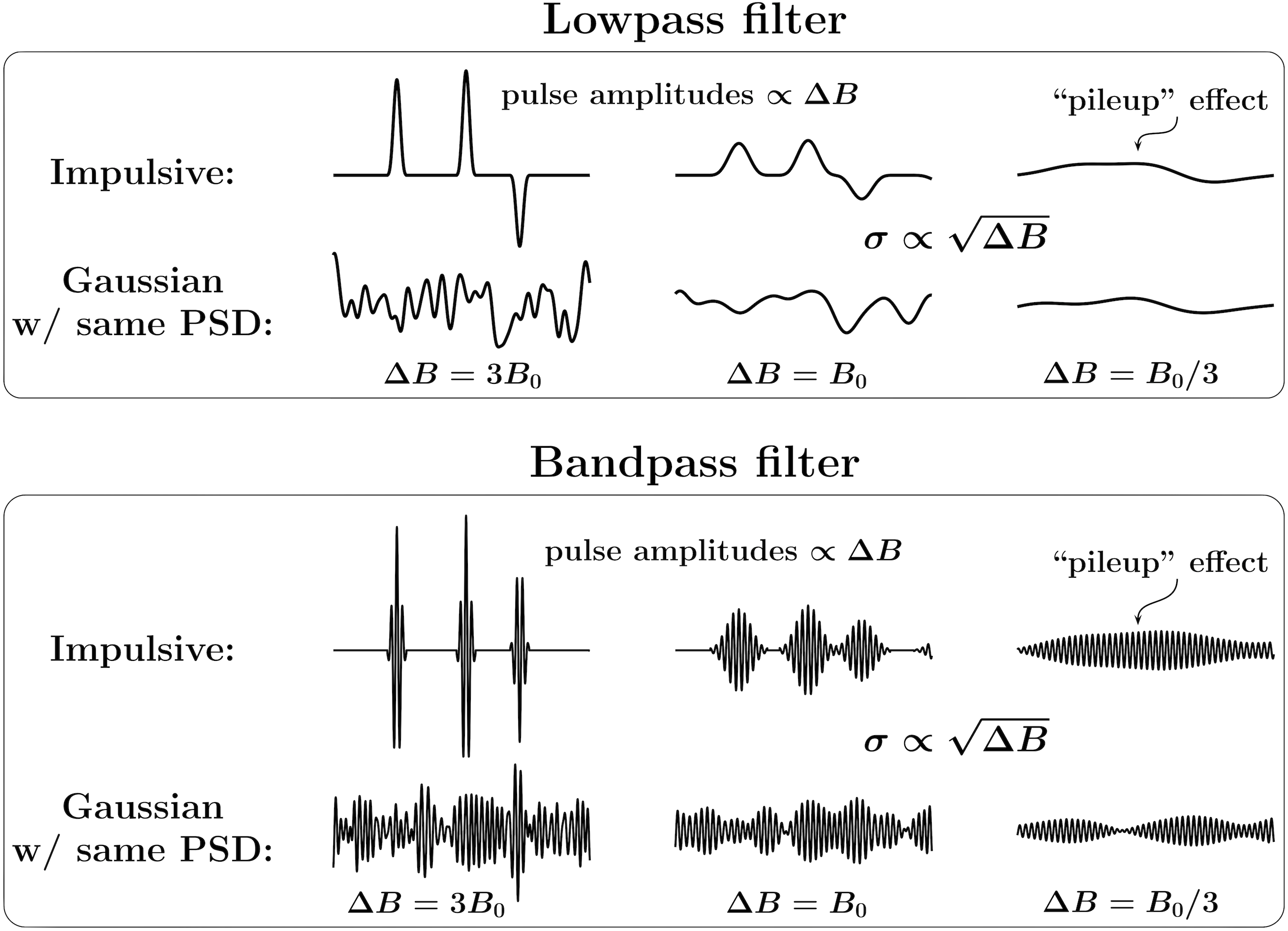}}
\caption{{\bf Reduction in bandwidth ``hides" outlier noise.}
Reproduced from~\cite{Nikitin19hidden}.
\label{fig:elusive}}
\end{figure}

Once outlier noise becomes apparent, additional reduction in bandwidth typically makes it less evident. Fig.~\ref{fig:elusive} illustrates the basic mechanism of outlier noise ``disappearance" with the reduction in observation bandwidth.

\begin{figure}[!t]
\centering{\includegraphics[width=8.6cm]{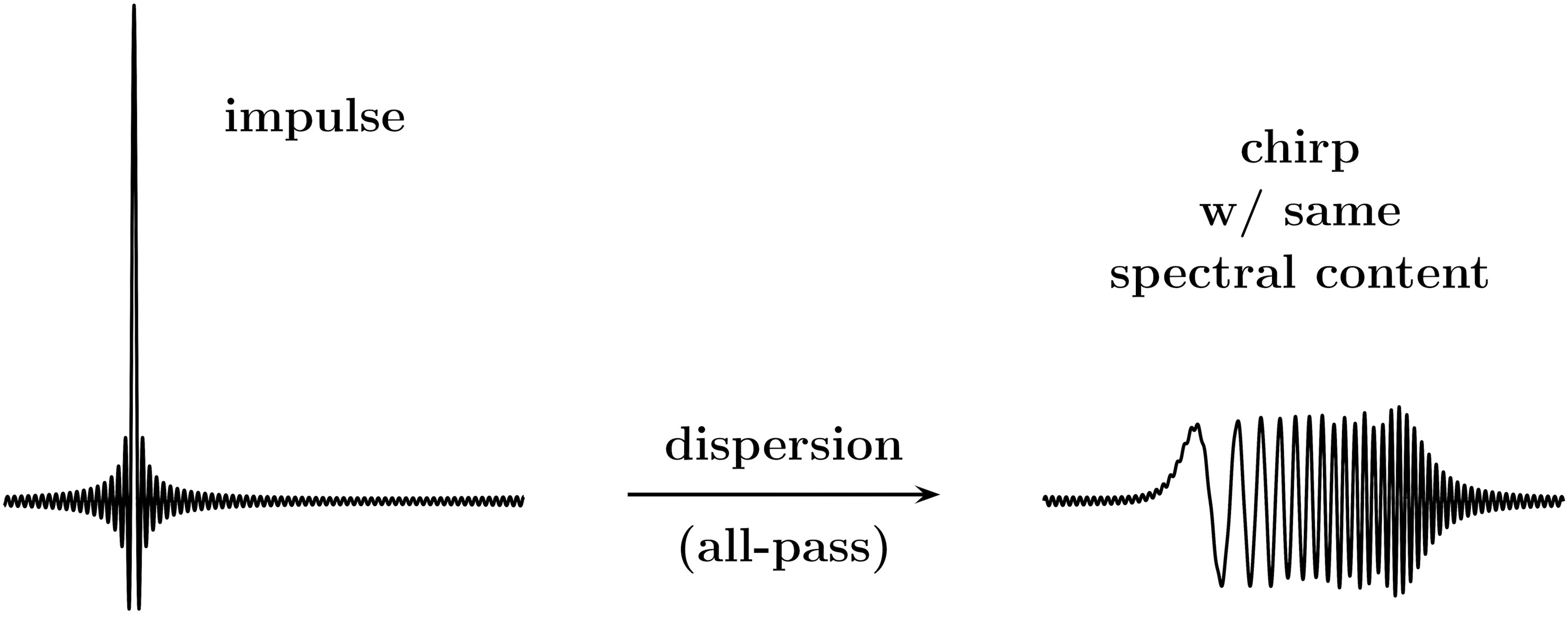}}
\caption{Impulse and chirp signals with same spectral content.
\label{fig:impulse2chirp}}
\end{figure}
\begin{figure}[!t]
\centering{\includegraphics[width=8.6cm]{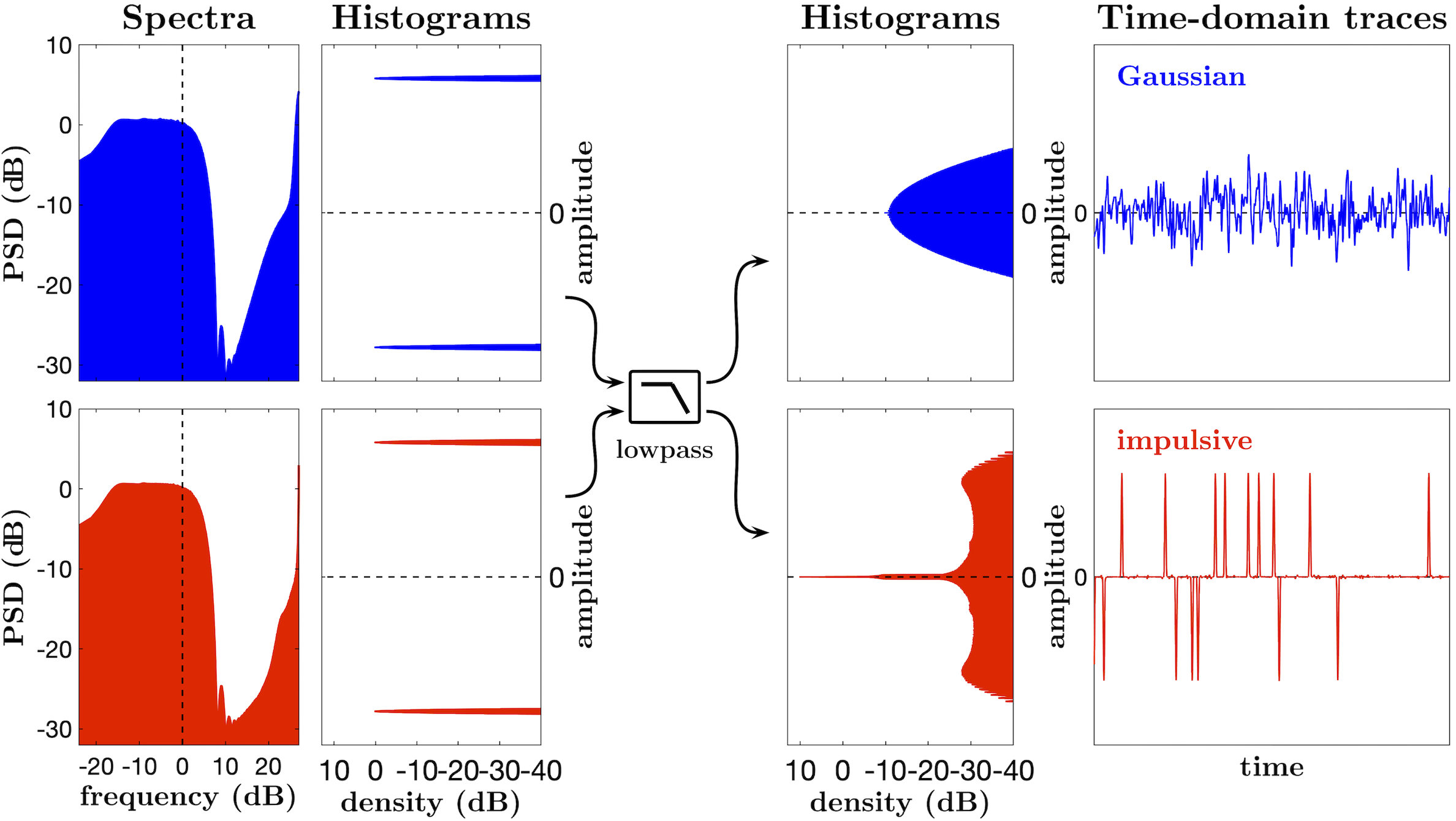}}
\caption{{\bf Conversely, reduction in bandwidth may reveal outlier noise.}
Here, original wideband signals are ``raw" outputs of 1-bit $\Delta\Sigma$ modulators given Gaussian or impulsive inputs.
\label{fig:DeltaSigma}}
\end{figure}

First note that a band-limited signal will not be affected by the change in the bandwidth of a filter, as long as the filter does not attenuate the signal's frequencies. Hence Fig.~\ref{fig:elusive} compares the effects of reducing the bandwidth of a lowpass or a bandpass filter only on Gaussian and impulsive noise. When the bandwidth~$\Delta{B}$ is reduced, the standard deviation of the noise decreases as a square root of its bandwidth, ${\sigma\propto\sqrt{\Delta{B}}}$. For Gaussian noise, its standard deviation is proportional to its amplitude. The amplitude of the impulsive noise, however, is affected differently by the bandwidth change. For example, as shown in Fig.~\ref{fig:elusive}, the amplitudes of the stand-alone pulses that effectively represent the impulse responses of the respective filters decrease proportionally to their bandwidth, faster than the amplitude of the Gaussian noise. Thus the bandwidth reduction causes the impulsive noise to protrude less from the Gaussian background. On the other hand, the width of the pulses is inversely proportional to the bandwidth. When the width of the pulses becomes greater than the distance between them, the pulses begin to overlap and interfere with each other. For a random pulse train, when the ratio of the bandwidth and the pulse arrival rate becomes significantly smaller than the time-bandwidth product of a filter, the resulting signal becomes effectively Gaussian due to the so-called ``pileup effect"~\cite[e.g.]{Nikitin98ppileup}, making the impulsive noise completely disappear.

In the tutorial presentation, we provide several detailed examples, along with the experimental evidence~\cite[e.g.]{Nikitin12aHSDPA}, of the effects of the observation bandwidth and the outlier occurrence rates on our ability to both observe and mitigate outlier noise.

\subsubsection{Spectral ambiguity} \label{subsubsec:spectral ambiguity}
Neither power spectral densities (PSDs) nor their short-time versions (e.g. spectrograms) allow us to reliably identify outliers, as signals with very distinct temporal and/or amplitude structures can have identical spectra. Fig.~\ref{fig:impulse2chirp} provides an emblematic example of such spectral ambiguity, depicting an impulse and a chirp signals with the same spectral content. This figure further illustrates that dispersion of a signal propagation medium may result in unintentional conversion of impulse signals into chirps. Thus filtering making the apparent noise outliers disappear and reappear does not have to affect the noise PSD.

\subsubsection{Ambiguity of amplitude densities} \label{subsubsec:density ambiguity}
Amplitude distributions (e.g. histograms) are also highly ambiguous as an outlier-detection tool. Although a super-Gaussian (heavy-tailed) amplitude distribution normally indicates presence of outliers, it does not necessarily reveal presence or absence of outlier noise in a signal+noise mixture. For example, two identical Gaussian mixture distributions can be composed of a narrowband Gaussian signal and a wideband impulsive noise, or a narrowband impulsive signal and a wideband Gaussian noise.

\subsubsection{Wide range of powers across spectrum} \label{subsubsec:power range}
Further, the outlier noise can be obscured by strong non-outlier signals, such as the thermal noise and/or adjacent channel interference, or by the signal of interest itself. More important, a wide range of powers across a wideband spectrum allows a signal containing outlier noise to have any type of amplitude distribution. This is illustrated in Fig.~\ref{fig:DeltaSigma}, where the original wideband signals are the ``raw" outputs of 1-bit $\Delta\Sigma$ modulators given a Gaussian or an impulsive inputs. Both are clearly two-level signals with identical sub-Gaussian amplitude distributions, and they may also represent bi-stable processes in general. However, in the same narrow band these signals have very different amplitude structures, one is Gaussian and the other one is super-Gaussian (impulsive).

\subsection{Outlier noise: Observation vs. mitigation} \label{subsec:observation}
We then proceed to show that, by utilizing a sufficiently wide ``excess band" and blocking those signals (e.g. signal of interest itself) that ``overpower" outlier noise, we can reliably identify such a noise in many cases when it would otherwise be obscured, therefore enabling its mitigation by intermittently nonlinear filters.

\subsection{Complex signal+noise compositions} \label{subsec:compositions}
We conclude Part~I of the presentation with a brief discussion of the general approach to the mitigation of interference that includes thermal noise, wideband outlier and non-outlier interference mixture, and that may also include a narrow-band outlier interference.
We emphasize that the wideband outlier noise must be mitigated {\em before\/} the final bandwidth reduction to that of the signal of interest, for example, in the process of the analog-to-digital conversion.

\section*{Part II: Methodology and tools for outlier noise mitigation} \label{sec:part II}
In Part~II, we describe the low-complexity intermittently nonlinear filtering arrangements that enable access to the ``untapped reserves" of interference mitigation. We then introduce the ``no harm" Complementary Intermittently Nonlinear Filtering (CINF) as a synergistic combination of linear and nonlinear filters, and provide quantitative illustrations of its performance. Subsequently, we discuss various approaches to CINF implementation, providing multiple examples of filtering configurations for addressing various practical interference scenarios.

\subsection{ADiC components and their implementation} \label{subsec:ADiC components}
First, we introduce the basic real-time nonlinear components that are used in outlier noise filtering, and give examples of their implementation. The main distinctive feature of these components is that they are {\em continuous-time\/} structures and thus can be implemented in {\em analog\/} circuitry. However, they can also be easily implemented numerically, e.g. in Field Programmable Gate Arrays (FPGA) or software. Such digital implementations would require little memory and would be typically inexpensive computationally (e.g. they are $\mathcal{O}(1)$ per output value in both time and storage), which makes them suitable for high-rate real-time digital processing.

\begin{figure}[!t]
\centering{\includegraphics[width=8.6cm]{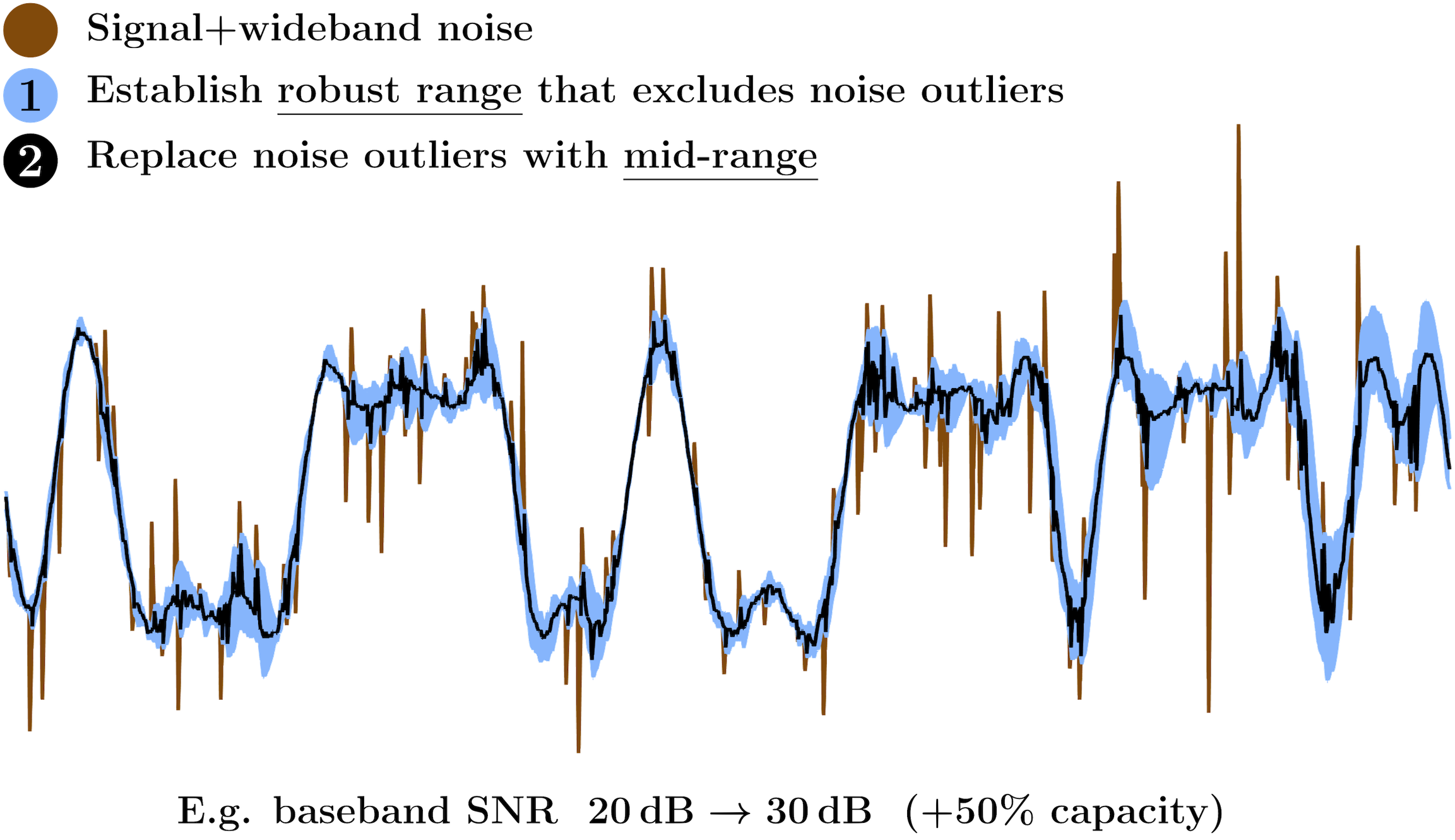}}
\caption{{\bf Removing wideband outlier noise while preserving signal of interest.}
Reproduced from~\cite{Nikitin19hidden}.\label{fig:ADiC concept}}
\end{figure}
\begin{figure}[!b]
\centering{\includegraphics[width=6.45cm]{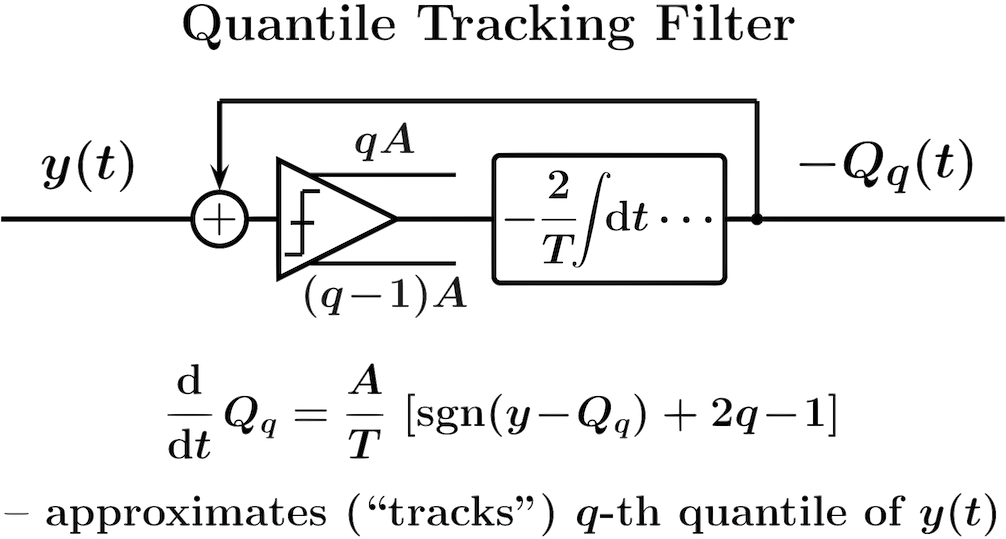}}
\caption{Block diagram of Quantile Tracking Filter (QTF).
\label{fig:QTF diagram}}
\end{figure}

\subsubsection{ADiC as main building block} \label{subsubsec:ADiC}
Fig.~\ref{fig:ADiC concept} illustrates the basic concept of an Analog Differential Clipper (ADiC) for wideband outlier noise removal while preserving the signal of interest and the wideband non-outlier noise. First, we establish a robust range that excludes noise outliers while including the signal of interest. Then, we replace the outlier values with those in mid-range. Note that this simplified illustration does not include any strong non-outlier wideband components, e.g. adjacent channel interference. Addressing such more complex interference compositions is discussed further in the tutorial.

\subsubsection{QTFs for robust range} \label{subsubsec:QTFs}
Fig.~\ref{fig:QTF diagram} shows a Quantile Tracking Filter (QTF) introduced in~\cite{Nikitin17nonlinear, Nikitin18ADiC-ICC} and described in detail in~\cite{Nikitin19ADiCpatentCIPs}. Given the input~$y(t)$, the output~$Q_q(t)$ of a QTF approximates (``tracks") the $q$-th~quantile of $y(t)$ obtained in a moving time window. (See~\cite{Nikitin03signal, Nikitin04adaptive} for discussion of quantiles of continuous signals.) Then various linear combinations of QTF outputs can be used to establish both a robust range ${[\alpha_-,\alpha_+]}$ that excludes outliers of a signal, and the mid-range that replaces the outlier values. For example, the range ${[\alpha_-,\alpha_+]}$ can be obtained as a range between {\it Tukey's fences\/}~\cite{Tukey77exploratory} constructed as linear combinations of the QTF outputs for the 1st ($Q_{[1]}$) and the 3rd ($Q_{[3]}$) quartiles:
\beginlabel{equation}{eq:Tukey's range}
  [\alpha_-,\alpha_+] = {\big [}Q_{[1]}\!-\!\beta\left(Q_{[3]}\!-\!Q_{[1]}\right)\!,\,Q_{[3]}\!+\!\beta\left(Q_{[3]}\!-\!Q_{[1]}\right)\!{\big ]},
\end{equation}
where $\beta$ is a scaling parameter of order unity (e.g. $\beta=1.5$).

\subsubsection{Basic ADiC structure} \label{subsubsec:basic ADiC}
In the basic ADiC structure the range is constructed as a range between Tukey's fences, and the mid-range is the arithmetic mean of these QTF fences (i.e. the arithmetic mean of the QTF outputs for the 1st and the 3rd quartiles).

\subsubsection{Much better way: Feedback-based ADiC} \label{subsubsec:feedback ADiC}
\begin{figure}[!t]
\centering{\includegraphics[width=8.6cm]{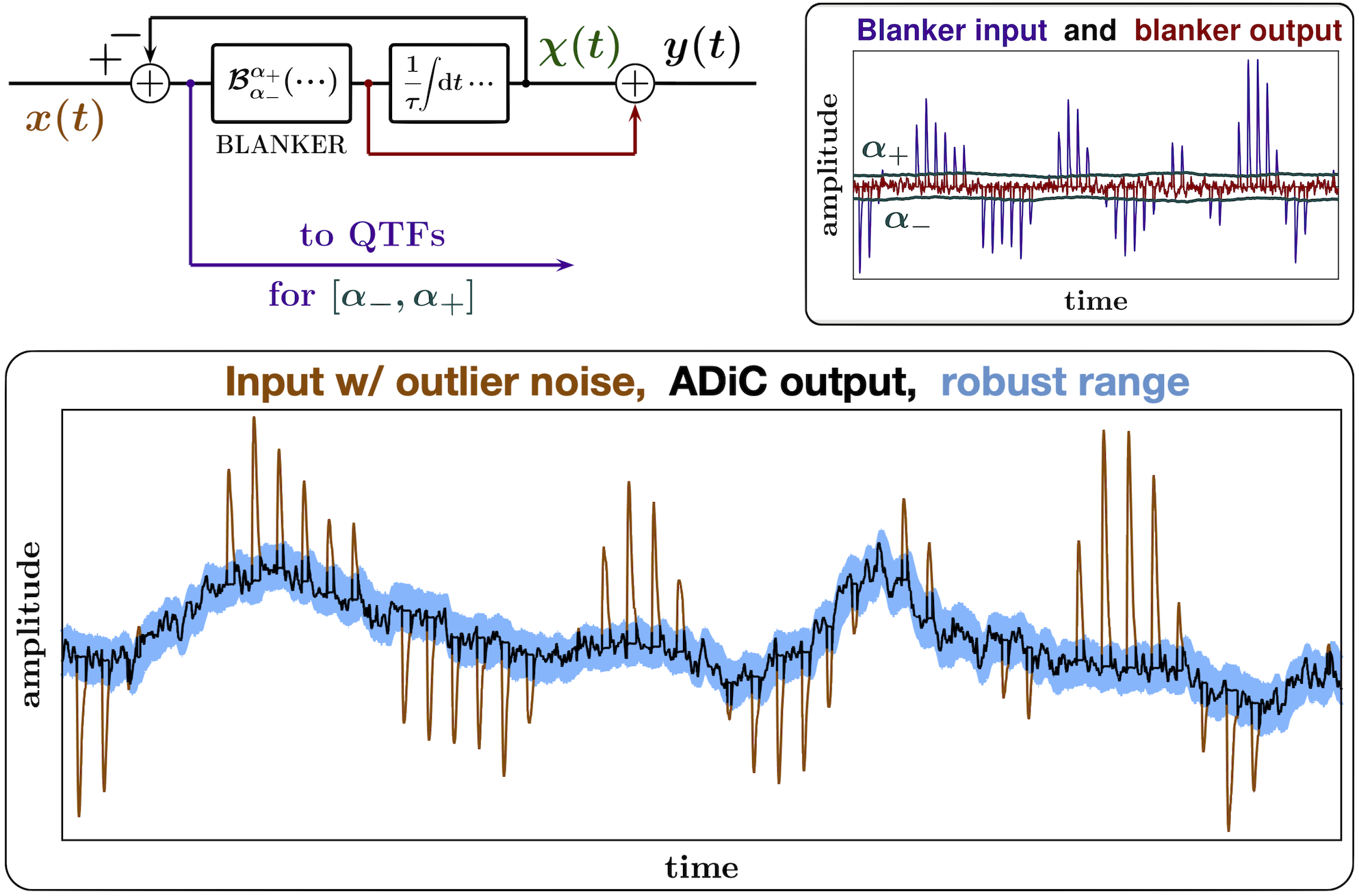}}
\caption{{\bf\boldmath Feedback-based ADiC replacing outliers with~$\chi(t)$.~}
Reproduced from~\cite{Nikitin19hidden}.
\label{fig:ADiC}}
\end{figure}

Fig.\,\ref{fig:ADiC} presents a feedback-based ADiC variant that has a number of practical advantages and is well suited for mitigation of hidden outlier noise~\cite{Nikitin19hidden}. As the diagram in the upper left of the figure shows, the ADiC output~$y(t)$ can be described as
\beginlabel{equation}{eq:ADiC equation}
  \left\{
  \begin{array}{cc}
    \!\! y(t) = \chi(t) + \tau\dot{\chi}(t)\\[1mm]
    \!\! \displaystyle \dot{\chi}(t) = \frac{1}{\tau}\, \BalphaPM\left(x(t)\!-\!\chi(t) \right)
  \end{array}\right.,
\end{equation}
where~$x(t)$ is the input signal, $\chi(t)$~is the {\em differential clipping level\/} (DCL), the {\em blanking function\/}~$\BalphaPM(x)$ is a particular type of an {\em influence function}~\cite{Hampel74influence} that is defined as
\beginlabel{equation}{eq:blanker}
  \BalphaPM(x)  = \left\{
  \begin{array}{cc}
    \!\! x & \mbox{for} \quad \alpha_- \le x \le \alpha_+\\
    \!\! 0 & \mbox{otherwise}
  \end{array}\right.,
\end{equation}
and where~$[\alpha_-,\alpha_+]$ is a robust range for the {\em difference signal\/}~${x(t)-\chi(t)}$ (the {\em blanking range\/}). Thus such an ADiC is an intermittently nonlinear filter that outputs the DCL~$\chi(t)$ only when outliers in the difference signal are detected, performing outlier noise mitigation without modifying the input signal otherwise. For the range fences such that ${\alpha_- \le x(t)\!-\!\chi(t) \le \alpha_+}$ for all~$t$, the DCL~$\chi(t)$ is the output of a 1st~order linear lowpass filter with the 3\,dB corner frequency~$1/(2\pi\tau)$. However, when an outlier of the difference signal is encountered, the rate of change of~${\chi(t)}$ is zero and the DCL maintains its previous value for the duration of the outlier.

\subsection{ADiC-based outlier noise filtering} \label{subsec:ADiC filtering}
In this section, we introduce Complementary Intermittently Nonlinear Filters (CINFs), and initially focus on the particular task of mitigating the outlier noise obscured by the signal of interest itself. Examples of particularly challenging waveforms that severely obscure low-amplitude outlier noise include broadband chirp signals used in radar, sonar, and spread-spectrum communications, and ``bursty," high crest factor signals such as those used in OFDM systems~\cite{Nikitin19complementary}.

\subsubsection{Spectral reshaping by ADiC and {\em efecto cucaracha}} \label{subsubsec:cucaracha}
We first point out that an ADiC applied to a filtered outlier noise can significantly reshape its spectrum. Such spectral reshaping by an ADiC can be called an {\em ``efecto cucaracha"\/} (a ``cockroach effect"), when reducing the effects of outlier noise in some spectral bands increases its PSD in the bands with previously low outlier noise PSD. We can use this property of an ADiC for removing outlier noise while preserving the signal of interest, and for addressing complex interference scenarios.

\subsubsection{CAF: Removing outlier noise while preserving signal of interest} \label{subsubsec:CAF}
\begin{figure*}[!t]
\centering{\includegraphics[width=16cm]{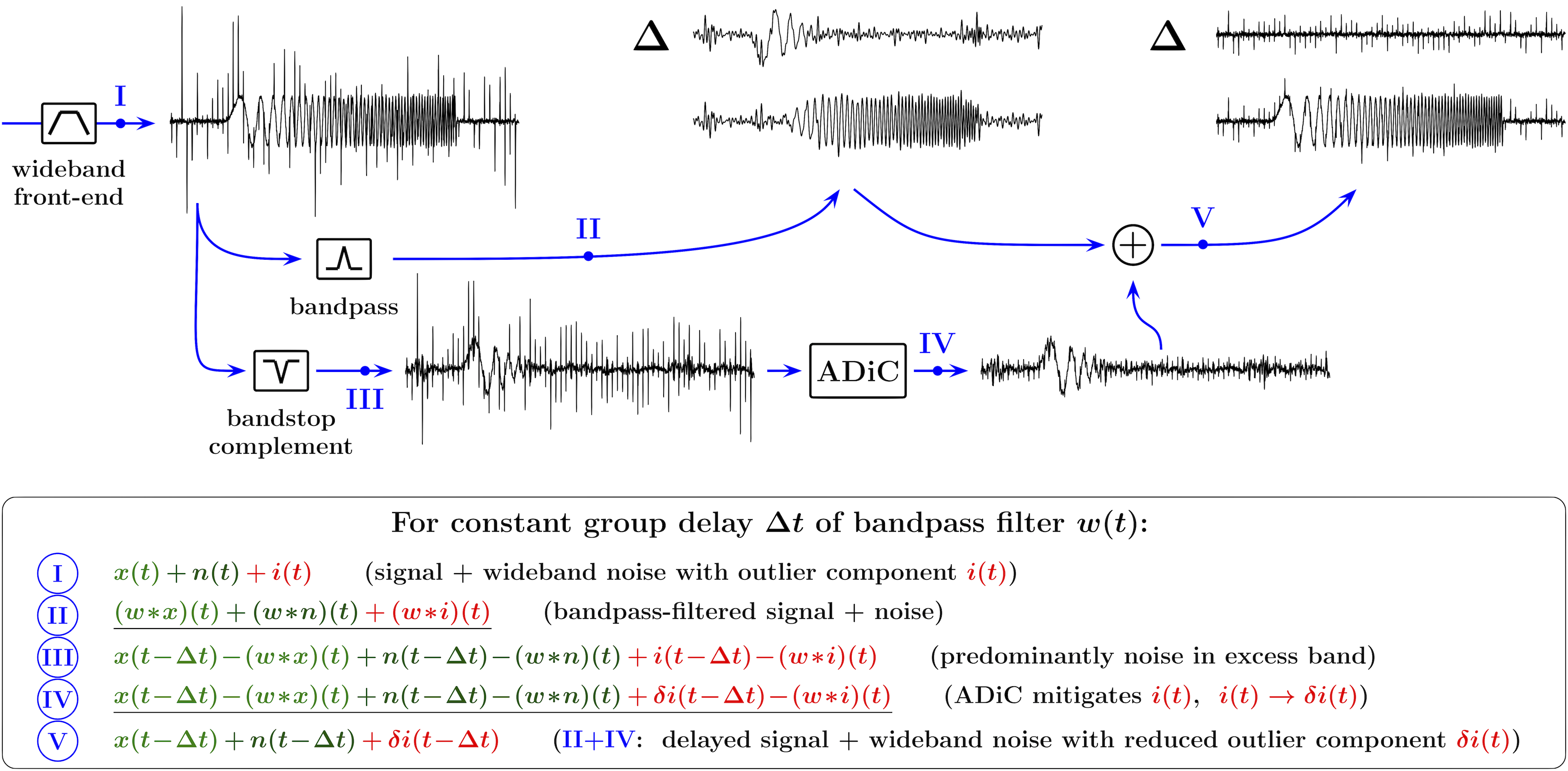}}
\caption{Complementary ADiC filtering (CAF) for removing wideband noise outliers while preserving band-limited signal of interest.
\label{fig:chirp CAF}}
\end{figure*}

For example, Fig.~\ref{fig:chirp CAF} illustrates the use of a Complementary ADiC Filtering (CAF) arrangement employed for mitigation of wideband outlier noise affecting a linear chirp signal. In Fig.~\ref{fig:chirp CAF}, the output~I of the wideband front-end filter consists of the chirp signal of interest~$x(t)$ and the wideband noise with non-outlier and outlier components~$n(t)$ and~$i(t)$, respectively. Due to high slew rates, the higher-frequency portion of the chirp signal is the most effective in obscuring low-amplitude broadband outliers. Therefore the bandstop complement of the bandpass filter should significantly reduce these high frequencies, in order for the outlier interference affecting the high-frequency portion of the chirp to become more conspicuous. On the other hand, the stopband of the bandstop filter should remain sufficiently narrow, so that the outlier in the combined impulse response of the front-end filter cascaded with the bandstop filter remains distinct. Thus, in this example the high-frequency edge~$f_{\rm c}$ of the bandstop filter is chosen at approximately the highest frequency of the chirp signal, and the low-frequency edge is placed at approximately~$f_{\rm c}/5$. While such a bandstop filter reduces the average slew rate of a linear chirp by about an order of magnitude, its stopband remains relatively narrow in comparison with the passband of the front-end filter, and the outlier structure of the noise is mainly preserved in the output~III of the bandstop filter. However, the bandstop filter significantly reduces the average slew rate of the chirp signal, making the outlier component~$i(t)$ more distinguishable and facilitating its efficient mitigation by an ADiC. As the result, the CAF output~V (that is the sum of the ADiC output~IV and the output~II of the bandpass filter) will consist of the effectively unmodified signal~$x(t)$ and the wideband noise with a reduced outlier component, $n(t)\!+\!\delta{i}(t)$, both delayed by the group delay~$\Delta{t}$ of the bandpass filter. In Fig.~\ref{fig:chirp CAF}, the traces marked by ``$\Delta$" show the respective differences between the filtered signal+noise mixtures and the delayed input signal~$x(t\!-\!\Delta{t})$ (without noise).

\subsubsection{CAF vs linear: Effect on channel capacity} \label{subsubsec:capacity}
We then outline the simulation setup for quantification of the improvements in signal quality provided by the CAF mitigation of outlier noise in comparison with linear filtering, and illustrate the relative increases in the baseband SNRs and in the channel capacities, as functions of the outlier-to-thermal noise power in the baseband, for different outlier noise compositions and moderate (10\,dB) and high (30\,dB) thermal noise SNRs.

\subsubsection{``No harm" (default) CAF configurations} \label{subsubsec:no harm}
In all these simulations, a ``default" set of CAF parameters was used, with the ``no harm" constraint such that nonlinear filtering does not degrade the resulting signal quality, as compared with the linear filtering, for any signal+noise mixtures. Thus, while providing resistance to outlier noise, in the absence of such noise CAFs behave effectively linearly, avoiding the detrimental effects, such as distortions and instabilities, often associated with nonlinear filtering. The ``no harm" property is especially important when considering complex, highly nonstationary interference scenarios, e.g.  in mobile and cognitive communication systems where the transmitter positions, powers, signal waveforms, and/or spectrum allocations vary dynamically. Note that when a CAF improves the signal quality, its performance can be further enhanced by optimizing its parameters.

\subsection{Analog vs digital} \label{subsec:analog vs digital}
The concept of ADiC-based filtering relies on continuous-time (analog) operations such as differentiation, antidifferentiation, and analog convolution. Therefore the most natural platform for implementing such filtering is analog circuitry. Analog processing is very appealing, e.g., when the requirements include inherently real-time operation, higher bandwidth, and lower power. On the other hand, digital processing offers simplified development and testing, configurability, and reproducibility. In addition, different ADiC components (e.g. QTFs) can be easily included into numerical algorithms without a need for separate circuits, and digital ADiC-based filtering is simpler to extend to complex-valued processing and to incorporate into various machine learning and optimization-based approaches.

\begin{figure}[!b]
\centering{\includegraphics[width=8.6cm]{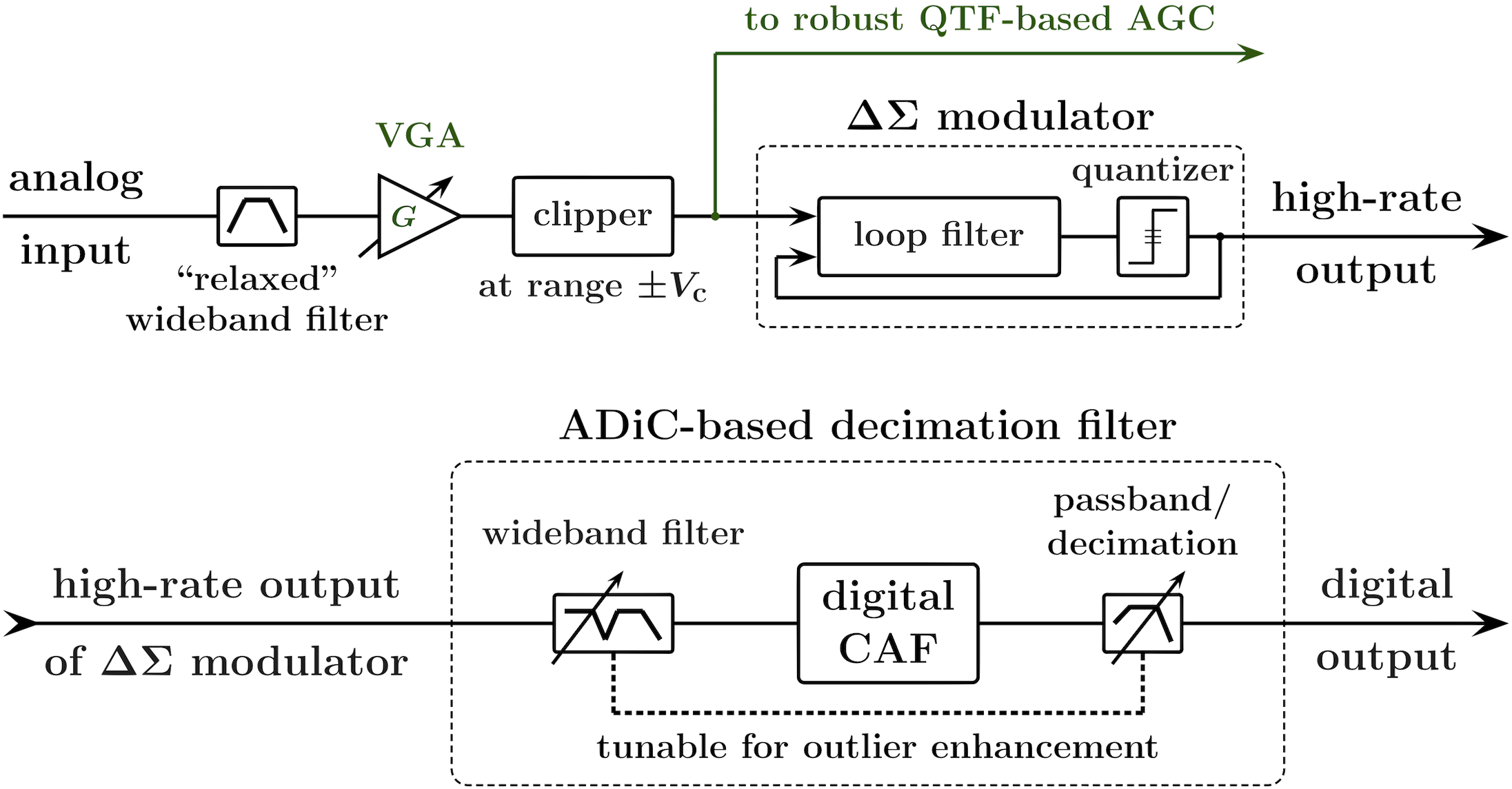}}
\caption{CAF in digital domain following $\Delta\Sigma$ modulator.
\label{fig:digital CAF}}
\end{figure}

\subsubsection{Digital: Where to get bandwidth?} \label{subsubsec:where bandwidth}
However, efficient mitigation of wideband outlier noise requires availability of a sufficiently broad excess band, and thus the respectively high ADC sampling rate. In addition, the sampling rate may need to be further increased so that analog differentiation can be replaced by its accurate finite-difference approximation, to enable ``effectively analog" processing. As described in our earlier work (see, e.g., \cite{Nikitin18ADiC-ICC, Nikitin19hidden, Nikitin19complementary}), we can use inherently high oversampling rate of a $\Delta\Sigma$~ADC to trade amplitude resolution for higher sampling rate and thus enable efficient digital ADiC-based filtering. Such an approach is illustrated in Fig.~\ref{fig:digital CAF}.

The high sampling rate allows the use of ``relaxed," wideband antialiasing filtering to ensure the availability of sufficiently wide excess band. As a practical matter, a wideband filter with a flat group delay and a small time-bandwidth product (e.g. with a Bessel response) should be used in order to increase the mitigable rates. Further, a simple analog clipper should be employed ahead of the $\Delta\Sigma$~modulator to limit the magnitude of excessively strong outliers in the input signal to the range~$\pm V_{\rm c}$ that is smaller than that of the quantizer. Such a clipper will prevent the modulator from saturation. In Fig.~\ref{fig:digital CAF}, a {\em robust\/} automatic gain control circuit (AGC) adjusts the gain~$G$ of the variable-gain amplifier (VGA) to maintain a constant output of a properly configured QTF circuit applied to the absolute value of the clipper output. This ensures that only large noise outliers are clipped, and not the outliers of the signal itself, e.g. outliers in high-crest-factor signals such as OFDM.

\subsubsection{Addressing complex interference scenarios} \label{subsubsec:complex}
Although it is perhaps unrealistic to require that any ``default" set of CINF parameters satisfies the ``no harm" constraint while improving the signal quality for all conceivable interference conditions, this constraint can always be met, for any particular interference scenario, in an ADiC-based filtering. For example, for an ADiC given by equation~(\ref{eq:ADiC equation}), in the limit of a wide blanking range such that ${\alpha_- \le x(t)\!-\!\chi(t) \le \alpha_+}$ for all~$t$, the ADiC becomes an allpass filter with a zero group delay, and it will not degrade the resulting signal quality, as compared with the linear filtering, for any signal+noise mixtures. Note, however, that when a CAF does improve the signal quality, its performance can be further enhanced by optimizing its parameters.

When observation bandwidth sufficiently larger than that of the signal of interest is available, various combinations of linear filters can be used to increase the difference between the temporal and/or amplitude structures of the interference and the signal of interest, enhancing the outlier components of the interference and enabling its mitigation by intermittently nonlinear filtering. Such combinations of front-end linear filters can be used not only for ``revealing" amplitude outliers hidden in broadband interference, but also for suppressing strong non-outlier interfering signals that may otherwise obscure these outliers, e.g. adjacent channel interference. Subsequently, if needed, the ``band of interest" filter (e.g. the digital decimation filter) can be modified to compensate for the impact of the front-end filter on the signal of interest. Several examples illustrating this technique can be found in~\cite{Nikitin19hidden, Nikitin19ADiCpatentCIPs}, and we discuss them in this tutorial. Specifically, we provide illustrations of mitigating sub-Gaussian outlier interference, and the outlier noise obscured by strong adjacent-channel interference.

\subsubsection{Practical configurations: CAF for chirp signals and OFDM} \label{subsubsec:chirp}
To emphasize the effectiveness and versatility of the CINF approach, we provide a detailed discussion of its practical application to particularly challenging waveforms that severely obscure low-amplitude outlier noise, such as broadband chirp signals (e.g. used in radar, sonar, and spread-spectrum communications) and ``bursty," high crest factor signals such as OFDM. We discuss co-design of the analog antialiasing and the digital pre-CAF filters, and describe a particular configuration of the QTF-based AGC circuit (see Fig.~\ref{fig:digital CAF}) tailored for OFDM signals. We also provide comparative simulation results for the filtering configuration shown in Fig.~\ref{fig:digital CAF}, for the cases of enabled and disabled ADiC-based filtering. 

\subsubsection{CAF for clipping distortions} \label{subsubsec:clipping distortions}
Further, we demonstrate the use of CAF for mitigation of clipping distortions as a particular type of outlier interference.

\subsubsection{Designing development \& testing platform} \label{subsubsec:PoC prototype}
\begin{figure*}[!t]
\centering{\includegraphics[width=15cm]{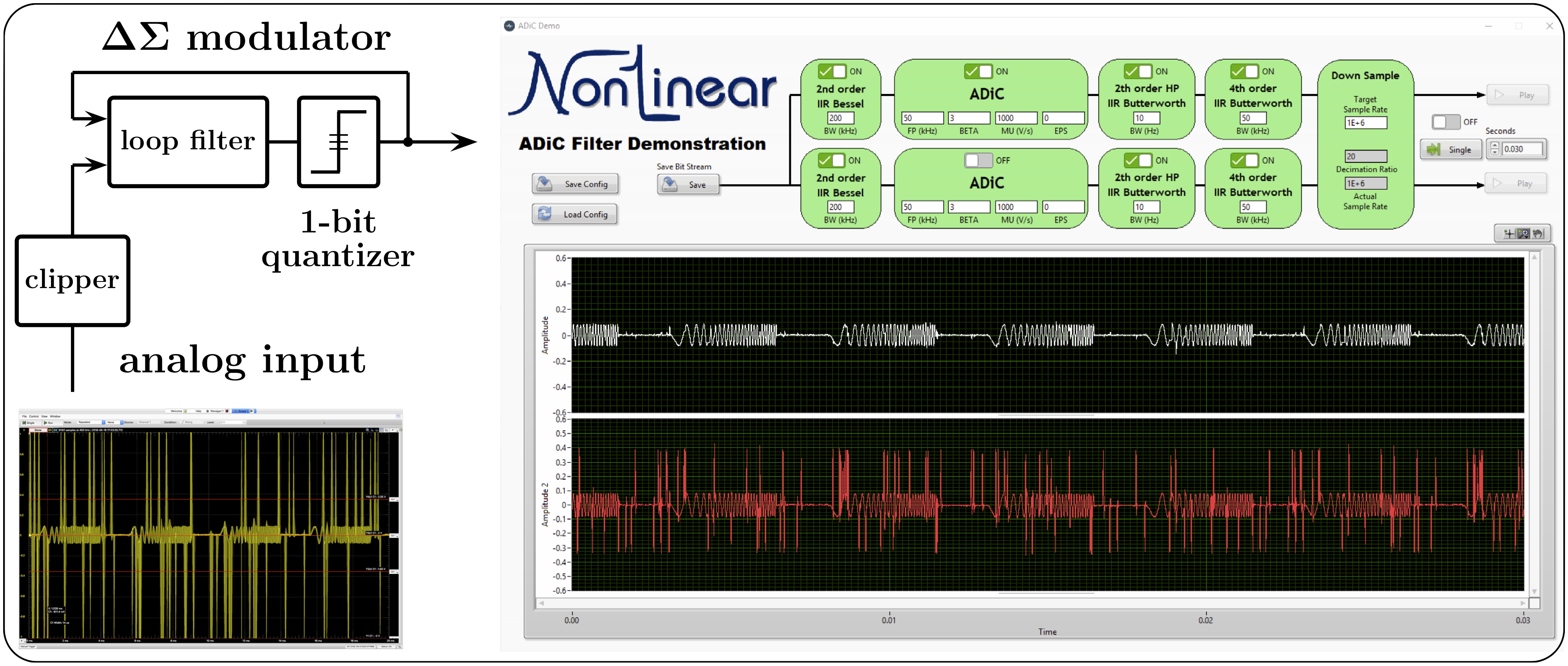}}
\caption{{\bf Prototype of ADiC filtering demo board.}
Reproduced from~\cite{Nikitin19hidden}.
\label{fig:demo}}
\end{figure*}

Fig.~\ref{fig:demo} shows an early prototype of an ADiC development and demonstration board that uses the ``effectively analog" implementation approach outlined above. This board employs the 1-bit isolated 2nd~order $\Delta\Sigma$ modulator AD7403, implements ADiC-based filtering in FPGA using National Instruments' (NI's) reconfigurable I/O (RIO) controller board NI sbRIO-9637, programmed using NI's LabVIEW graphical development environment and LabVIEW FPGA module. This allows fast and easy reconfigurability of the ADiC-based processing for evaluating the performance of alternative ADiC topologies and their dependence on the ADiC parameters. In addition to testing and displaying the comparative results of the ADiC-based filtering for various waveforms and noise compositions, in the frequency range for up to several hundreds of kilohertz, this board allows real-time audible range demonstrations with instant playback. This development board is a step toward application-specific ADiC configurations.

\subsection{Broader picture} \label{subsec:broader picture}
\begin{figure}[!b]
\centering{\includegraphics[width=8.6cm]{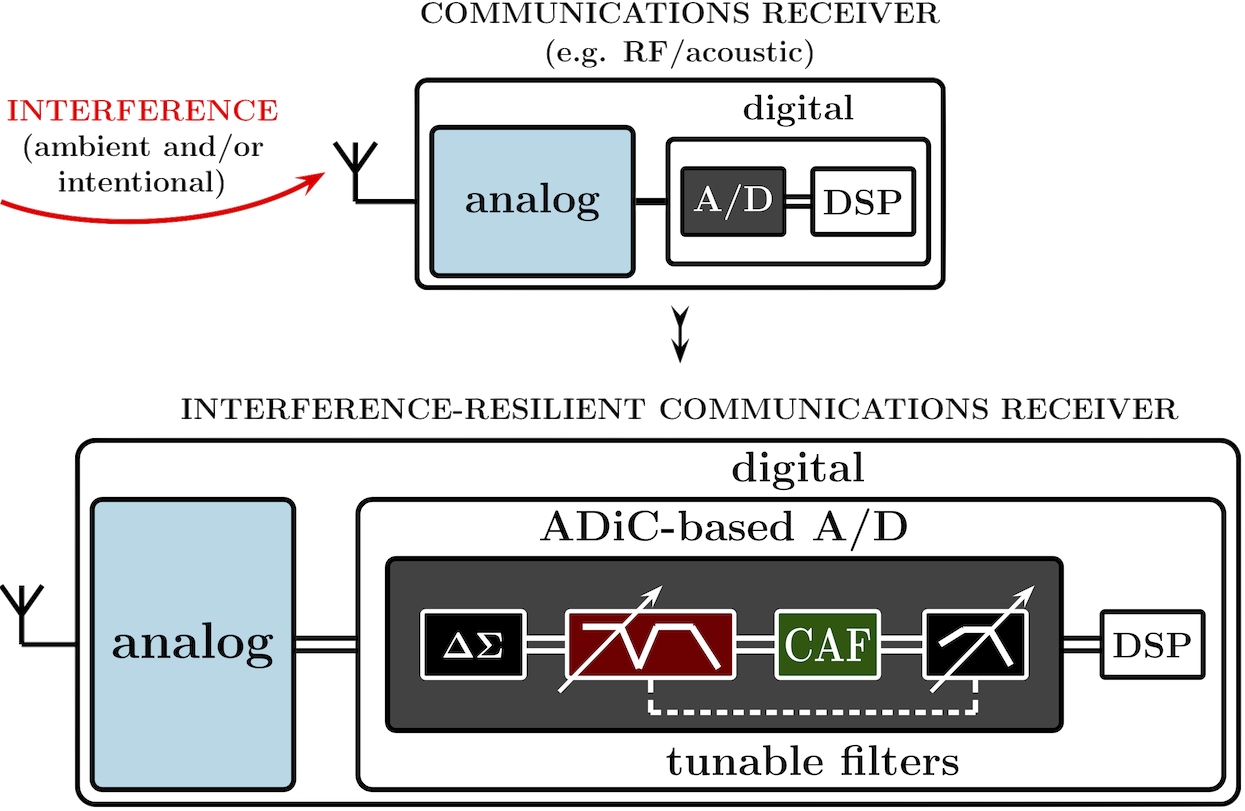}}
\caption{{\bf Interference resilient ADiC-based receiver.}
Reproduced from~\cite{Nikitin19hidden}.
\label{fig:InterferenceResilient}}
\end{figure}

Fig.~\ref{fig:InterferenceResilient} summarizes the potential use of the ADiC-based A/D conversion for development of communication receivers resilient to outlier interference of various types and origins, including those due to intermodulation distortions (IMD) and spectral regrowth caused by strong signals. This approach can be integrated into existing communication systems, e.g. implemented with existing communication radios operating in the HF, VHF and UHF spectrum. As discussed above, tunable wideband linear filters can be deployed ahead of the CAF for outlier enhancement, and various machine learning and optimization-based techniques can be used for their tuning to optimize the receiver performance for particular system configurations and/or interference scenarios.

\section*{Conclusion} \label{sec:conclusion}
In this tutorial, we provide an overview of the methodology and tools for real-time mitigation of outlier interference in general and ``hidden" wideband outlier noise in particular. Either used by itself, or in combination with subsequent mitigation techniques, this approach provides interference reduction levels otherwise unattainable, with the effects, depending on particular interference scenarios, ranging from ``no harm" to considerable. While the main focus of this filtering technique is to serve as a ``first line of defense" against wideband interference ahead of, or in the process of, analog-to-digital conversion, it can also be used, given some {\em a priori} knowledge of the signal of interest's structure, to reduce outlier interference that is confined to the signal's band.

In addition to addressing ``hidden interference" scenarios, the distinct feature of the proposed approach is that it capitalizes on the ``excess band" observation of interference for its efficient in-band mitigation by intermittently nonlinear filters. This significantly extends the mitigation range, in terms of both the rates of the outlier events and the mitigable SNRs, in comparison with the mitigation techniques focused on the apparent in-band effects of outlier interference.

The CINF-based structures described in the tutorial are mostly ``blind" as they do not rely on any assumptions for the underlying interference beyond its ``inherent" outlier structure, yet they are adaptable to nonstationary signal and noise conditions and to various complex signal and interference mixtures. Thus they can be successfully used to suppress interference from diverse sources, including the RF co-site interference and the platform noise generated by on-board digital circuits, clocks, buses, and switching power supplies. They can also help to address multiple spectrum sharing and coexistence applications (e.g. radar-communications, radar-radar, narrowband/UWB, etc.), including those in dual function systems (e.g. when using radar and communications as mutual signals of opportunity). This filtering paradigm can further benefit various other military, scientific, industrial, and consumer systems such as sensor/sensor networks and coherent imaging systems, sonar and underwater acoustic communications, auditory tactical communications, radiation detection, powerline communications, navigation and time-of-arrival techniques, and many others. Finally, various embodiments of the presented filtering structures can be integrated into, and manufactured as IC components for use in different products, e.g. as A/D converters with incorporated interference suppression.

\section*{Acknowledgment}
The authors would like to thank
Keith~W. Cunningham of Atkinson Aeronautics \& Technology Inc., Fredericksburg, VA;
Scott~C. Geier of Pyvonics LLC, Garden City, KS;
James~E. Gilley of BK Technologies, West Melbourne, FL;
William~B. Kuhn of Kansas State University, Manhattan, KS;
Earl McCune of Eridan Communications, Santa Clara, CA;
Alexey~A. Nikitin of AWS, Seattle, WA;
Arlie Stonestreet\,\,II of Ultra Electronics ICE, Manhattan, KS,
and Kyle~D. Tidball of Textron Aviation, Wichita, KS,
for their valuable suggestions and critical comments.
This work was supported in part by Pizzi Inc., Denton, TX 76205 USA.


\end{document}